# Application of time-reversal-based processing techniques to enhance detection of GPR targets

Vinicius Rafael N. Santos[1] and Fernando L. Teixeira

Electroscience Laboratory and Dept. of Electrical and Computer Engineering

The Ohio State University, 1330 Kinnear Rd., Columbus, OH 43212, USA

[1]nerisdossantos.1@osu.edu

## Abstract

In this paper we analyze the performance of time-reversal (TR) techniques in conjunction with various Ground Penetrating Radar (GPR) pre-processing methods aimed at improving detection of subsurface targets. TR techniques were first developed for ultrasound applications and, by exploiting the invariance of the wave equation under time reversal, can yield features such as superresolution and statistical stability. The TR method was examined here using both synthetic and actual GPR field data under four different pre-processing strategies on the raw data, namely: mean background removal, eigenvalue background removal, a sliding-window space-frequency technique, and a noise-robust spatial differentiator along the scan direction. Depending on the acquisition mode, it was possible to determine with good precision the position and depth of the studied targets as well as, in some cases, to differentiate the targets from nearby clutter such as localized geological anomalies. The proposed methodology has the potential to be useful when applied to GPR data for the detection of buried targets.

# 1. Introduction

Ground penetrating radar (GPR) is widely used for mapping and identification of buried targets, such as concrete tubes, metallic and plastic drums, pipes, etc. However, conventional data processing methods for target detection do not always yield satisfactory results in field applications due to the presence of ground clutter, ground losses, and geological anomalies that often obscure the target response and create ambiguities. In a typical GPR deployment, a transmitter (or set of transmitters) located above ground sends a waveform into the subsurface and the resulting scattered signal by the target(s) and geological anomalies within the subsurface is collected by a receiver (or set of receivers) above ground. Since the wave equation governing propagation of the waveforms is invariant under time-reversal (TR) in stationary and lossless media and approximately so in low-loss media, this implies that the data collected by the receivers can be time-reversed and used to recreate wavefields that would "backpropagate" into the subsurface and automatically focus on reflective targets and/or anomalies[1]. In general, TR invariance can be exploited for detection and localization of obscured targets in noisy and rich-scattering environments, with high resolution. Time-reversal (TR) techniques were first developed for acoustics (Fink et al., 1989) (Fink, 1992), being later successfully used in several applications including non-destructive testing and evaluation (Liu et al., 2014), sound quality enhancement (Lin & Too, 2014), atmospherics studies (Mora et al., 2012), subsurface geophysics (Fink, 2006; Leuschen & Plumb, 2001; Saillard et al., 2004; Cresp et al., 2008; Artman et al., 2010; Foroozan & Asif, 2010; Yavuz et al., 2014; Chen et al., 2016), microwave remote sensing (Reyes-Rodríguez et al., 2014), wireless communications (Fouda et al., 2012) and medicine (Thomas & Fink, 1996; Tanter et al., 1996). The basic TR algorithm can be modified in a number of ways; for example, it can be applied iteratively in sequence to automatically focus on the most reflective target immersed in a scattering medium (Prada et al., 1993). Variants on the basic TR algorithm exist which allow for selective focusing on well-resolved secondary (weaker) targets (Saillard et al., 2004; Yavuz & Teixeira, 2006) and tracking of obscured moving targets as well (Fouda & Teixeira, 2012).

---

[1] Backpropagation can be effected either physically by transmitting the time-reversed signals into the original medium, or synthetically by means of a forward simulation engine, such as the finite-difference time-domain (FDTD) method. Synthetic backpropagation is done for imaging purposes, as considered here.

In this paper, we compare the results of the TR algorithm when combined with four preprocessing techniques for GPR data, to determine the most effective combination thereof: Background removal (Daniels, 2007), eigenvalues background removal (Shahbaz & Al-Nuaimy, 2009), a sliding-window space-frequency technique (Yavuz et al., 2014) and a noise-robust spatial differentiator along the scan direction (Holoborodko, 2016). The comparative analysis is performed using both simulation results based on the Finite-Difference Time-Domain (FDTD) algorithm and field data from actual measurements. Consideration of actual field data allows for assessing the performance under typical signal-noise ratio found in practice. Consideration of simulation data allows for a better control and subsequent assessment of the variables affecting the algorithm performance. Both types of data can suggest ways to improve these methods in practice. The results obtained here can provide tools for improving GPR data interpretation and may assist in future applications of GPR in geotechnical and environmental studies.

## 2. TR technique: Fundamentals

To make the discussion here more self-contained, we next briefly describe the basic principles of the TR technique. More details can be found in Fink (2006). Although it has been developed originally for acoustic waves, the TR technique can be equally applied to electromagnetic waves (Yavuz & Teixeira, 2009). This process is akin to Reverse Time Migration (RTM) utilized for seismic imaging (Baysal et al., 1983). The standard TR technique comprises four basic steps and a scheme is shown in Figure 1: (1) A pulse signal is transmitted from one or more transceivers into the scattering medium under interrogation; (2) the transmitted signal propagates through the scattering medium and it is reflected by one or more targets; (3) The reflected signals are collected by the transceivers (Figure 1a); (4) Each of the signals thus collected are reversed in time and subsequently sent back (in a first-in last-out fashion) by the corresponding transceivers to the same medium again (Figure 1b).

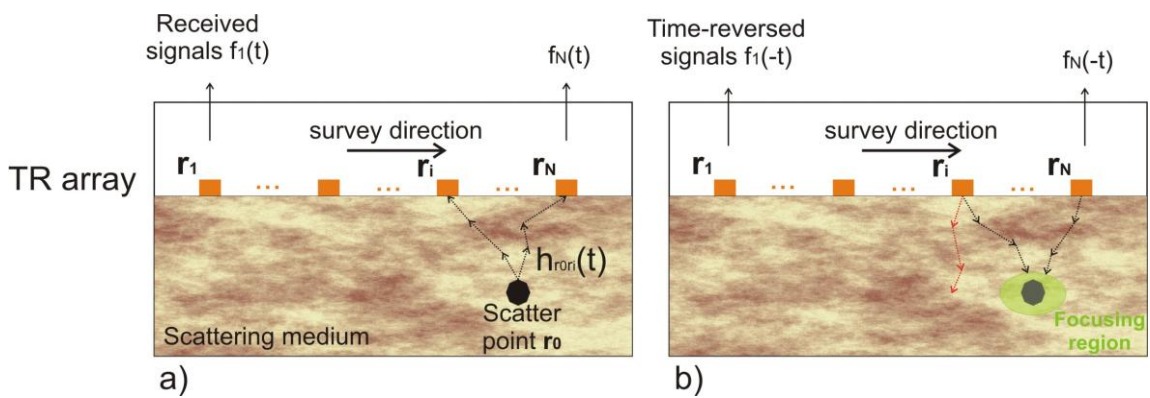

**Figure 1.** Basic time-reversal procedure. a) Physical scattering from a buried target after the forward propagation of the GPR signal. b) Synthetic back-propagation of time-reversed signals.

The TR invariance of the wave equation implies that, in stationary and low-loss media, the retransmitted signals will automatically focus around the original source (target) of the scattered signal. To further understand this fact, note that the solution $\vec{E}(\vec{r},t)$ of wave equation for the electric field in a lossless medium obeys

$$\nabla^2 \vec{E}(\vec{r},t) - \mu(\vec{r})\varepsilon(\vec{r})\frac{\partial^2}{\partial t^2}\vec{E}(\vec{r},t) = 0 \tag{1}$$

where $\vec{r}$ is the position vector, and μ and ε are the permeability and permittivity of the medium, respectively (Yavuz & Teixeira, 2009). It is clear that this equation also admits $\vec{E}(\vec{r},-t)$ as solution. Assuming that a transmitter antenna located at $\vec{r_0}$ sends a pulse *s(t)*, the signal collected by some receiver antenna located at $\vec{r_i}$ can be expressed as the convolution

$$f_i(t) = s(t) * h(t) \tag{2}$$

where *h(t)* is the channel (impulse) response between the antennas $\vec{r_0}$ and $\vec{r_i}$. To stress the spatial dependency, the channel response can be expressed as $h_{\vec{r_0}\vec{r_i}}(t)$. From the reciprocity theorem (Chew, 1994) it follows that

$$h_{\vec{r_0}\vec{r_i}}(t) = h_{\vec{r_i}\vec{r_0}}(t) \tag{3}$$

Mathematically, the TR operation consists of transmitting a time-reversed copy of (2) by the antenna at location $\vec{r_i}$, that is

$$f_i(-t) = s(-t) * h_{\vec{r_0}\vec{r_i}}(-t) \tag{4}$$

As a result, the time-reversal signal obtained at the original antenna location $\vec{r_0}$ becomes

$$p_i(t) = s(-t) * h_{\vec{r_0}\vec{r_i}}(-t) * h_{\vec{r_i}\vec{r_0}}(t) = s(-t) * h_{\vec{r_0}\vec{r_i}}(-t) * h_{\vec{r_0}\vec{r_i}}(t) \tag{5}$$

which can be recognized as a *matched filter* operation in both space and time, *regardless of the specific channel response $h_{\vec{r_0}\vec{r_i}}(t)$*. This matched filter operation yields both spatial focusing and temporal compression. Somewhat surprisingly, both these features are *enhanced* in rich-scattering environments (Yavuz & Teixeira, 2005).

More generally, multiple transmitter and receiver antennas can be used, comprising what is called a time reversal array (TRA). Assuming a TRA with *N* elements (transceivers), the received signal is given by

$$p(\vec{r_0}, t) = \sum_{i=1}^{N} s(-t) * h_{\vec{r_0}\vec{r_i}}(-t) * h_{\vec{r_i}\vec{r_0}}(t) \tag{6}$$

To illustrate this process, Figure 2a shows an example of radargram obtained for a plastic pipe buried in a homogeneous soil, where the reflection time is 7.85 ns. The TR result can be encoded into a three-dimensional (3D) matrix, with the three index sets representing (1) the horizontal distance along the acquisition path, (2) the vertical depth, and (3) the time sample (time step index), respectively. This 3-D data set can also be encoded as a set of 2-D matrices *M*[*x,y*], with each matrix representing a time-step, the result of the first TR set correspond to all the first samples for each A-scan and so on until the number of samples. Figures 2b to 2j show nine time-steps of the TR result applied to the plastic pipe data. At successive iterations, the wavefield tends to focus near the position of the target (pipe). As seen in Figure 2f, at about 7.85 ns the peak of the wavefield distribution coincides with the horizontal pipe position.

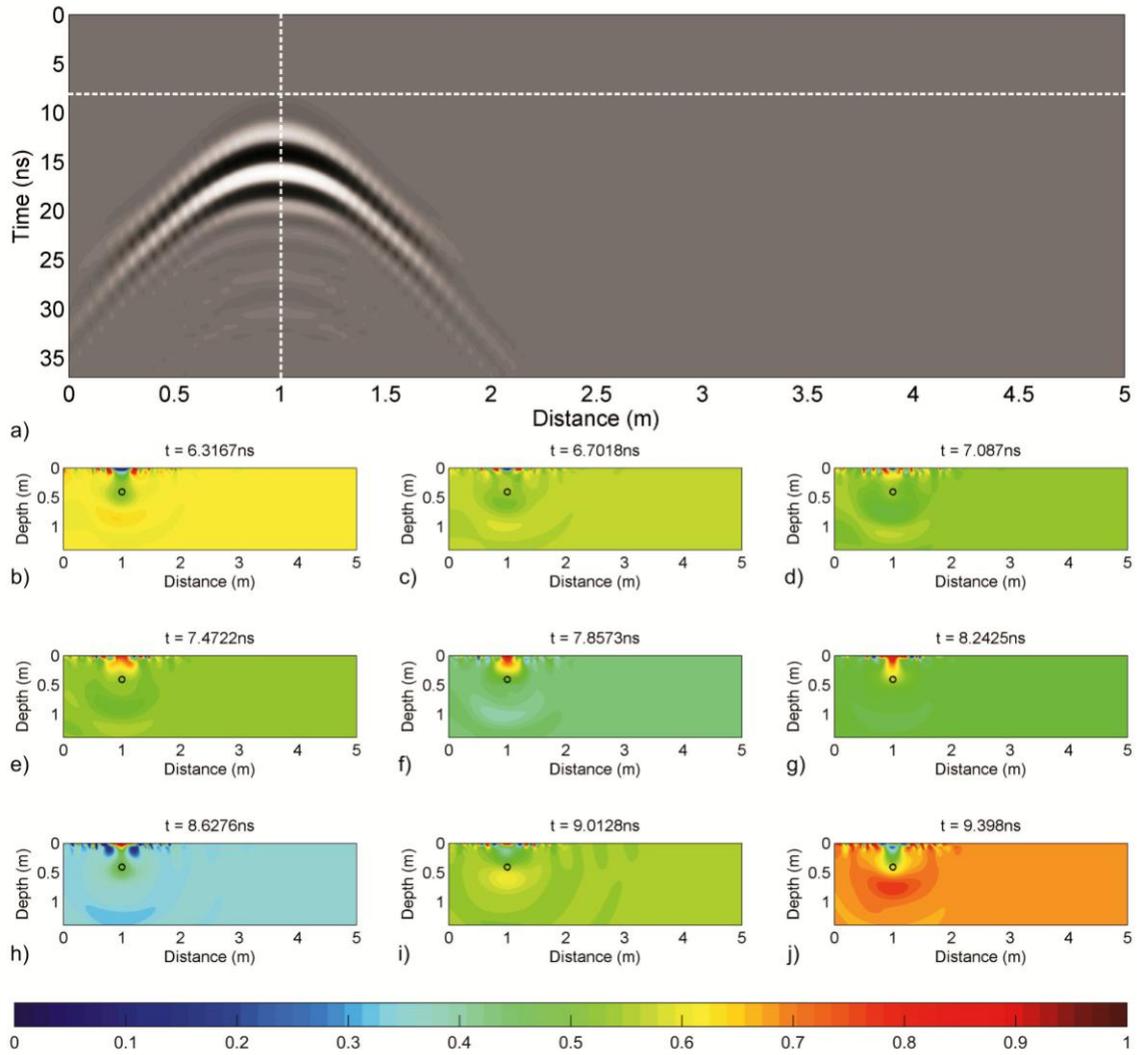

**Figure 2.** a) Synthetic GPR data for buried plastic pipe. b) to j) Different time-steps from TR results applied to GPR data.

## 3. Data processing

One way to perform the TR analysis is to directly analyze the time evolution of the backpropagated signal, as shown on Figure 2. Potential scatter locations would then correspond to regions where localized focusing is observed in the domain. However, since this focusing occurs only during specific time intervals, there is an inherent ambiguity in the interpretation of the results unless the time intervals of significant focusing are known a priori, which is not the case in practice. Furthermore, in practice there is no precise information about the pointwise permittivity of the soil, which causes a mismatch between the subsoil medium assumed during synthetic backpropagation and the actual subsoil where the radargram is obtained. This mismatch produces both spatial and temporal blurring of the focusing of synthetic fields, which can further confound data interpretation.

To overcome these issues, we propose here to use of the standard deviation of the amplitude of the backpropagated TR wavefield, sampled either along time or space (depth or co-range). This standard deviation provides the dispersion of each data set in such a way that larger range of variation in the TR amplitudes caused by focusing effects are emphasized independently from of the estimation of precise focusing intervals a priori. The standard deviation is computed for the whole data set (three-dimensional TR matrix). The basic sequence of steps is shown in the diagram of Figure 3. The first step consists of pre-processing of the GPR using one of many available techniques, with four possible ones indicated in Figure 3.

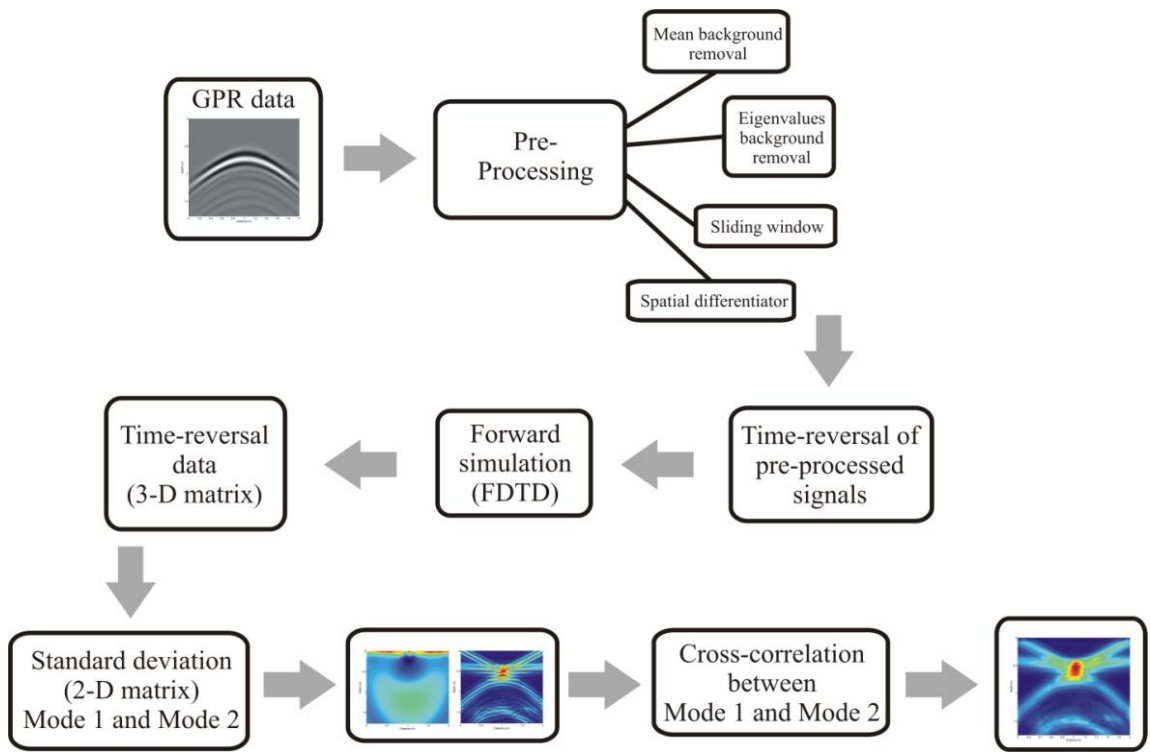

**Figure 3.** Diagram with the basic steps utilized for GPR data processing.

Different pre-processing techniques highlight different aspects of the data. Time-reversal is next employed to the pre-processed data, i.e. each A-scan time-domain signal is reversed in a first-in, last-out fashion. These reversed signals are used a waveform excitations on the GPR antenna and their backpropagation into the domain is simulated using a 3-D FDTD code. This simulation produces a so-called time-reversal data, in the form of a three-dimensional matrix. We consider two strategies here for computing standard deviations from the time-reversal data, denoted as Mode 1 and Mode 2:

(a) *Mode 1*: This mode is summarized in Fig. 4. The set of all A-scans for each time sample produces a corresponding TR matrix. Figure 4a shows the GPR data where each horizontal line (six are shown) represents a set of A-scans that leads to a set of TR matrices (Figure 4b). The standard deviation is computed between the samples of every TR data set on the same position across the TR matrices in Figure 4b. For example, the four colored points in each of the matrices of Figure 4b indicate four positions where the standard deviation was computed, with result shown in Figure 4c. By considering more points, we can determine the variation curve along each TR depth samples, one of them shown in Figure 4c. Finally, by using all A-scans, it is possible to produce an image of the standard deviation along the A-scan-depth axis as shown in Figure 4d. The horizontal line on Figure 4d indicates the target location along the track.

(b) *Mode 2*: This mode is summarized in Figure 4. As before, each time sample of the radargram data (Figure 5a) produces a corresponding TR matrix. In this mode, the standard deviation is computed for each A-scan of each TR matrix. For example, Figure 5b shows the TR matrix corresponding to the 350th time sample (time step sequence in the FDTD simulation). The horizontal lines in Figure 5b indicate different set of samples from which the standard deviation is computed in this mode. Note that the same TR matrix produces a set of standard deviation along different A-scans, as shown in Figure 5c. Finally, by considering all TR matrices (different time samples) the standard deviation image shown in Figure 5d can be generated. The yellow line in Figure 5d marks the position of the target along the track.

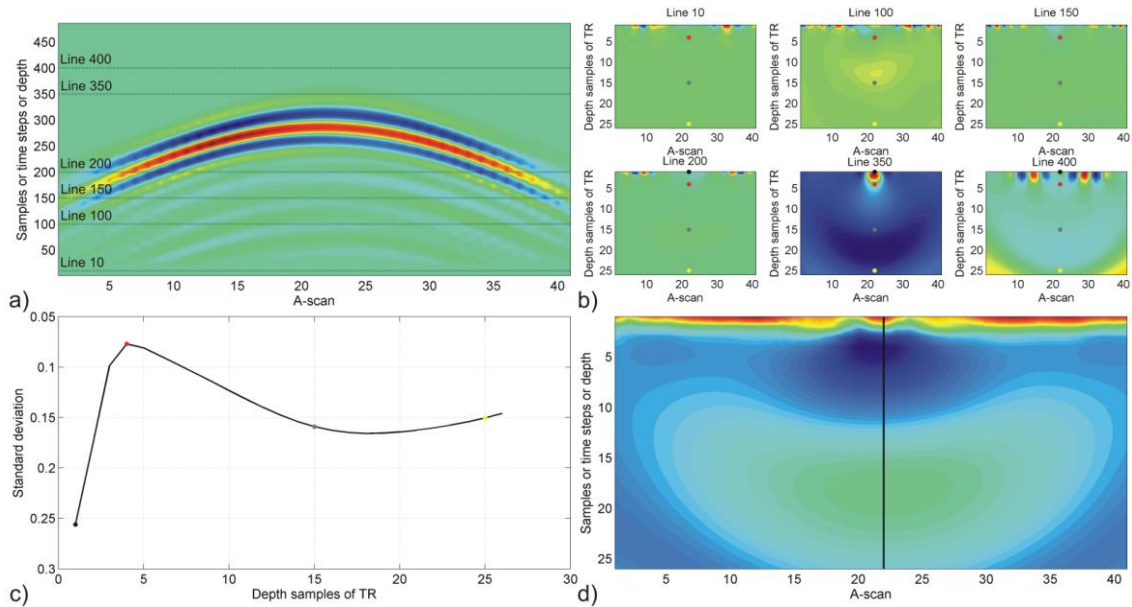

**Figure 4.** Standard deviation computations for Mode 1. a) GPR data. b) Time-reversal data in different time steps. c) Standard deviation of A-scan 22. d) Data set of standard deviation TR.

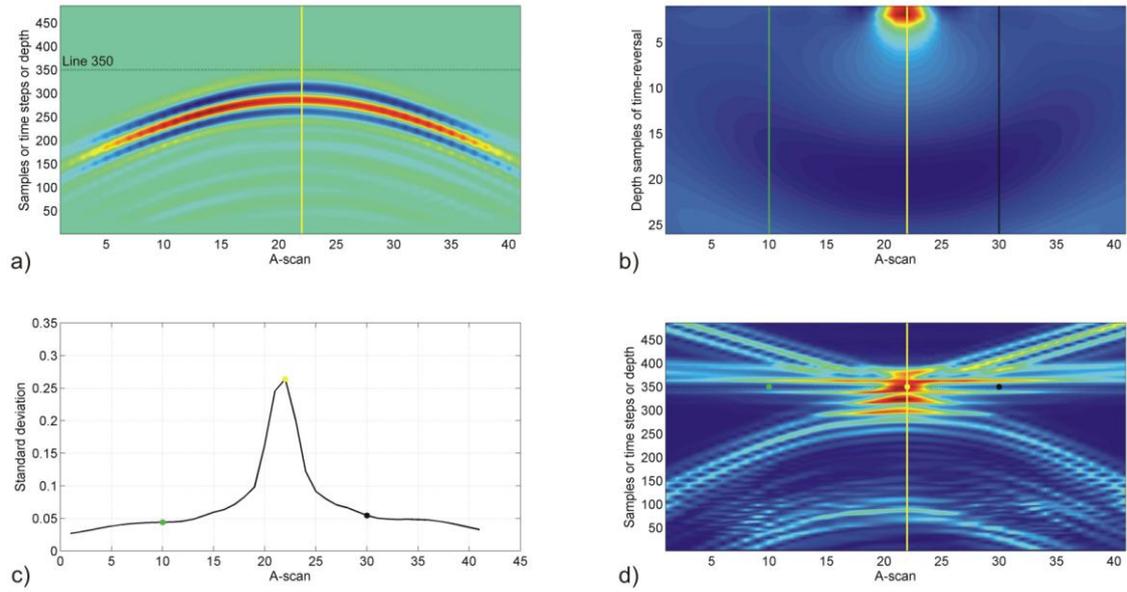

**Figure 5.** Standard deviation computations for Mode 2. a) GPR data. b) Time-reversal data in time step 350. c) Standard deviation of A-scan 22. d) Data set of standard deviation TR.

Mode 1 tends to emphasize the target position along the GPR track, whereas Mode 2 tends to emphasize the target depth. Although data from Mode 1 and Mode 2 can in principle be utilized and interpreted separately, they are susceptible to spurious artifacts such as "ringing" effects that may confuse the interpretation. In order to mitigate image artifacts and improve the results, we also present results based on the cross-correlation between Mode 1 and Mode 2, which we denote Mode X12. This measures the similarity of the data sets. Mode X12 does not provide added information but it emphasizes common features of both modes, greatly facilitating interpretation.

We analyzed the raw data obtained from the FDTD modeling as well after different types of pre-processing in order to verify which combination of pre-processing and standard deviation modes result provides good results. As noted before, four different pre-processing techniques were considered for inter-comparison: (1) mean background removal, (2) eigenvalues background removal, (3) sliding-window space-frequency technique, and (4) along-scan spatial derivative.

## 4. Simulation setup and data pre-processing techniques

To obtain synthetic radargrams, we utilize a forward simulations based on a 3-D FDTD algorithm with perfectly matched layers (PML) (Teixeira et al., 1998; Lee et al., 2004). The central frequency used in the simulations was 200 MHz and the background soil properties were assumed that of a homogeneous medium with relative permittivity of 15 and conductivity of 3 mS/m. Two different models were built to simulate targets commonly found on geotechnical studies as shown schematically in Figure 6. The first model is depicted in Figure 6a and includes a plastic pipe (ø = 0.1 m) and a vertical metallic drum (ø = 0.65 m and 0.9 m length), at positions 1 m and 3.5 m from the origin of the profile and at 0.4 m and 0.5 m depths, respectively. The second model is depicted in Figure 6b and has the same plastic pipe from the first model, at the same position and depth, plus a metallic pipe (ø = 0.2 m), at position 3.5 m and at 0.5 m of depth, and a small (ø = 0.2 m) geological anomaly. The geological anomaly exhibit a fluctuating relative permittivity with values between 15±3. The latter model seeks to simulate the response of geologic anomalies that can be confused with that of real targets and a target that has permittivity very close to that of the background soil, configuring a more challenging detection problem.

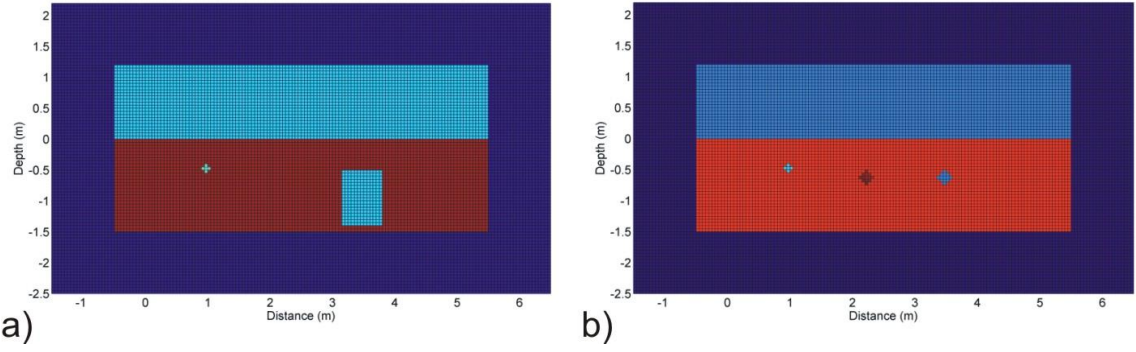

**Figure 6.** Models used in this study. a) Plastic pipe and vertical metallic drum. b) Horizontal plastic pipe and metallic pipe, and punctual geological anomaly.

From the FDTD results, the next step was the application of the TR method to the numerical data. For that, it is necessary to reverse in time all signals obtained from the radargram, and transmitting them back into to studied medium, using the FDTD code. To reduce the redundancy on the data set and highlight the anomalies, the TR signal amplitudes are rescaled between the range [0,1], i.e.

$$\tilde{x}_i = \frac{X_i - X_{min}}{X_{max} - X_{min}} \qquad (7)$$

where $x_i$ is each sample amplitude, $x_{min}$ is the minimum value and $x_{max}$ is the maximum value of the set.

The raw data obtained by the FDTD algorithm is shown in Figures 7a and 7b, for Model 1 and Model 2, respectively. The profiles have 5 m of length and 1.4 m of depth. The geological anomaly was detected, but its response is quite similar to that of the pipe response, confounding detection. The only difference between these two responses is the signal amplitude, which results from the permittivity contrast, closer to the medium.

The first pre-processing applied in to the raw data was mean background removal. This is a widely used pre-processing technique utilized for GPR radargrams, and it can be expressed as (Daniels, 2007):

$$A'_{n,a}(t) = A_{n,a}(t) - \frac{1}{N_a}\sum_{a=1}^{N_a} A_{n,a}(t) \tag{8}$$

where $A_{n,a}(t)$ is the unprocessed A-scan and $A'_{n,a}(t)$ is the processed A-scan, $N_a$ is the number of A-scan waveforms, and the above is repeated for $n = 1$ to $N$, where $N$ is the number of sample. This pre-processing suppresses contributions from horizontal layers, and importantly from the air-ground interface. The results of the mean background removal can be seen in Figures 7c and 7d. Note that some inevitable horizontal artifacts are produced in the image, a consequence of "spreading" of the waveform mean in eq. (8).

The next signal pre-processing technique is the eigenvalue-based method developed for GPR scenarios by Khan and Al-Nuaimy (2009). This technique exploits the fact that some eigenvalues of a B-scan are related to the target response, while some eigenvalues are related to noise in the data (Cagnoli and Ulrych, 2001). A given B-scan can be represented in terms of its Singular Value Decomposition (SVD) as follows:

$$X = \sum_{i=1}^{r} \sigma_i \mu_i \vartheta_i^T \tag{10}$$

where $X$ is the B-scan matrix, $r$ is the rank of $X$, $\mu_i$ is the $i^{th}$ eigenvector of $XX^T$, $\vartheta_i$ is the $i^{th}$ eigenvector of $X^T X$, $\sigma_i$ is the $i^{th}$ singular value of $X$. The aim of this method is to determine the eigenvalue set that provide maximum suppression of clutter noise while keeping the signal from the targets of interest. Results from the application of eigenvalues background removal are shown in Figures 7e and 7f. The ground-antenna coupling was removed from data and the shape of the hyperbolas was preserved. The bottom of the metallic drum was detected (7e) and the geologic punctual anomaly was removed during the processing (7f).

The sliding-window space-frequency technique was first developed by Yavuz et al. (2014). In this technique, the submatrices (sub-B-scan) that form a given radargram are utilized to extract localized scattering information. Each sub-B-scan is decomposed into its singular vectors and weighted by the singular values subtracting from the full B-scan, thereby reducing the clutter and enhanced target response.

The submatrix corresponding to a $M$ x $L$ sub-B-scan is written as

$$B_i^{sub} = B_i^{sub}(R_i^{L+i-1}, \omega) = \begin{pmatrix} A(\mathbf{r}_i, \omega_1) & \cdots & A(\mathbf{r}_{L+i-1}, \omega_1) \\ \vdots & \ddots & \vdots \\ A(\mathbf{r}_i, \omega_M) & \cdots & A(\mathbf{r}_{L+i-1}, \omega_M) \end{pmatrix} \quad (11)$$

where $i$ is the number of A-scans considered in a single sub-B-scan. By applying a singular value decomposition to each $B_i^{sub}$, it results

$$B_i^{sub} = U_i^{sub} \Lambda_i^{sub} (V_i^{sub})^* \quad (12)$$

where $U_i^{sub}$ is the unitary $M$ x $M$ matrix of left singular vectors, $V_i^{sub}$ is the unitary $L$ x $L$ matrix of right singular vectors, and $\Lambda_i^{sub}$ is the $M$ x $L$ matrix containing the singular values. Here, the superscript * denotes conjugate transpose operator.

The results of the application of sliding-window are shown in Figures 7g and 7h. All anomalies can be detected with this technique. The boundaries of vertical target are well delimited as shown in Figure 7g.

The last pre-processing technique considered is a smooth noise-robust differentiator of the radargram amplitude trace along the scan direction (Holoborodko, 2016). The equation that describes this processing is:

$$A'_a(t) = -\frac{1}{8}A_{a-2}(t) - \frac{2}{8}A_{a-1}(t) + 0A_{a+0}(t) + \frac{2}{8}A_{a+1}(t) + \frac{1}{8}A_{a+2}(t) \quad (13)$$

with $a = 1$ to $N_a$, and $N_a$ is the number of A-scan waveforms. In the above, $A_{n,a}(t)$ is the raw unprocessed A-scan, and $A'_{n,a}(t)$ is the processed A-scan. The result of this processing is displayed in Figures 7i and 7j, where it is possible to see all anomalies present.

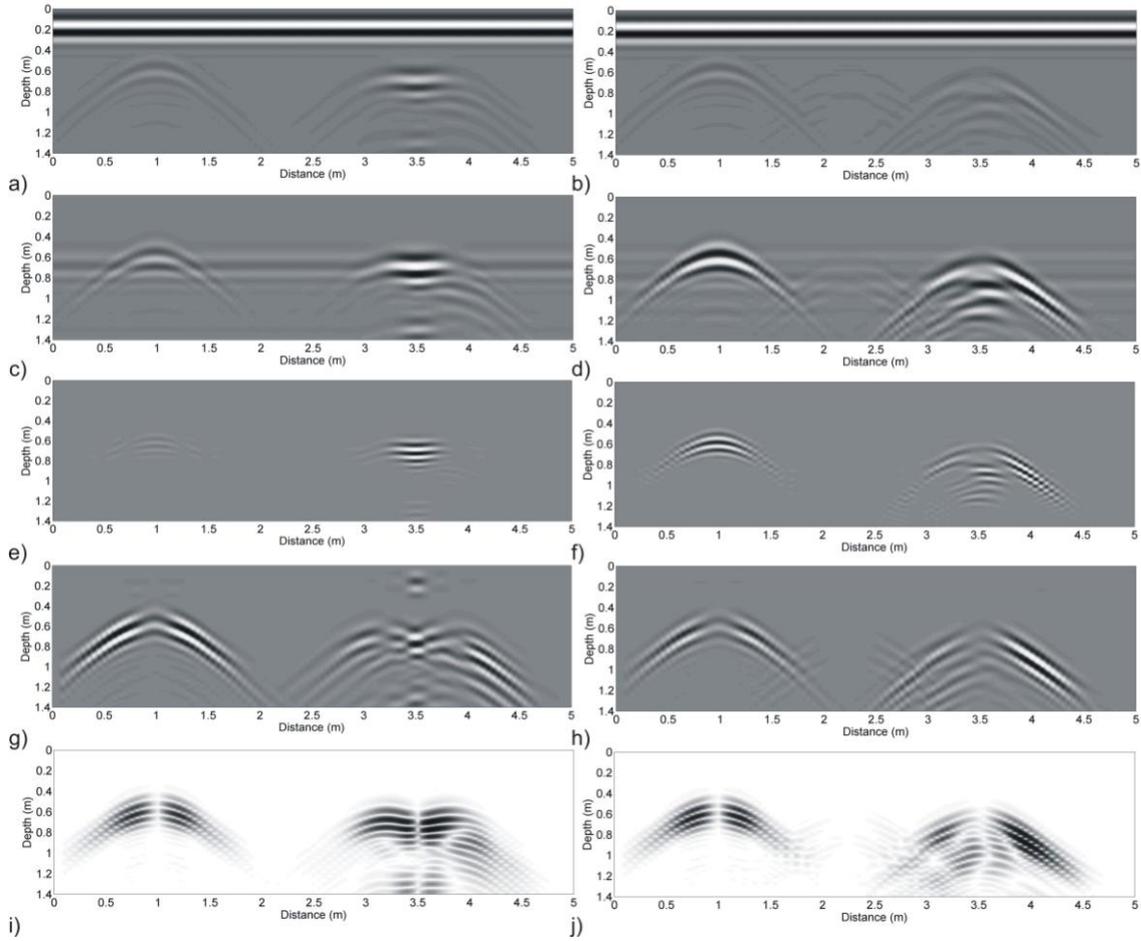

**Figure 7.** Synthetic GPR data. First column: Model 1. Second column: Model 2. a) and b) Raw data. c) and d) Mean background removal data. e) and f) Eigenvalues background removal data. g) and g) Sliding-window data. i) and j) Along-track spatial derivative processing data.

## 5. Simulation results

Figure 8 show the application of TR with Mode 1, Mode 2 and their cross-correlation, Mode X12. The circles and rectangle drawn with black lines indicate the position of the pipes and drum, respectively, and the white lines indicate the position of the geological anomaly. Using Mode 1, it is difficult to detect the presence of the discrete targets and only the metallic drum (Figure 8a) is easily discernible. Limited resolution along track exists, but, as expected, depth information is not retrieved. On the other hand, Mode 2 determines both the position and depth of the targets more precisely (Figures 8b and 8e). In particular, the response from the metallic drum exhibits a very good lateral resolution, and the punctual geologic anomaly can be visibly discerned. Appling Mode X12 smoothes the data (Figures 8c and 8f) and suppresses image spurious artifacts. In this case, it is possible to better discern all anomalies present, including the lateral limits of the metallic drum again and a better focusing in the pipe positions.

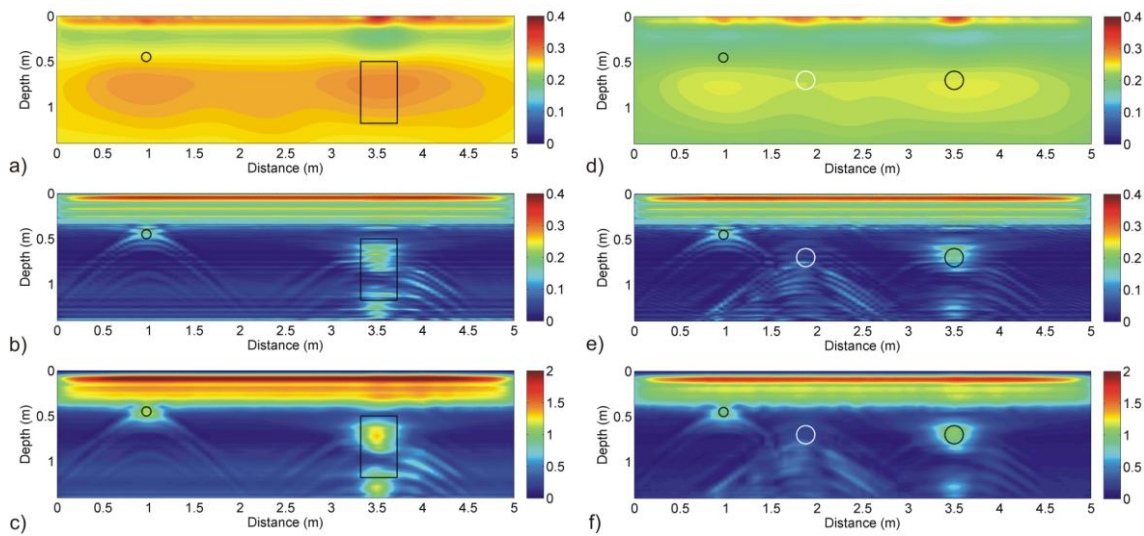

**Figure 8.** TR results for Models 1 and 2, applied to the raw GPR data. a) and d) Mode 1. b) and e) Mode 2. c) and f) Mode X12.

Figure 9 shows the results of the TR for Modes 1 (9a and 9d), 2 (9b and 9e) and X12 (9c and 9f) after the mean background removal. The horizontal reflection was removed from the data and the anomalies are highlighted in Mode 1 (differently of the raw data), showing the presence of targets. The resolution along track is improved with respect to the raw data but, again, depth resolution is not retrievable in Mode 1. Under Mode 2, better along-track resolution is obtained. Depth information is also obtained; however, some artifacts remain present. Importantly, Figures 9e and 9f show that the response from the geological anomaly is minimized, with potential for facilitating target discrimination. Mode X12 better highlights the anomalies, suppressing the ringing artifacts visible in Mode 2. The metallic pipe anomaly has approximately the same size as the pipe.

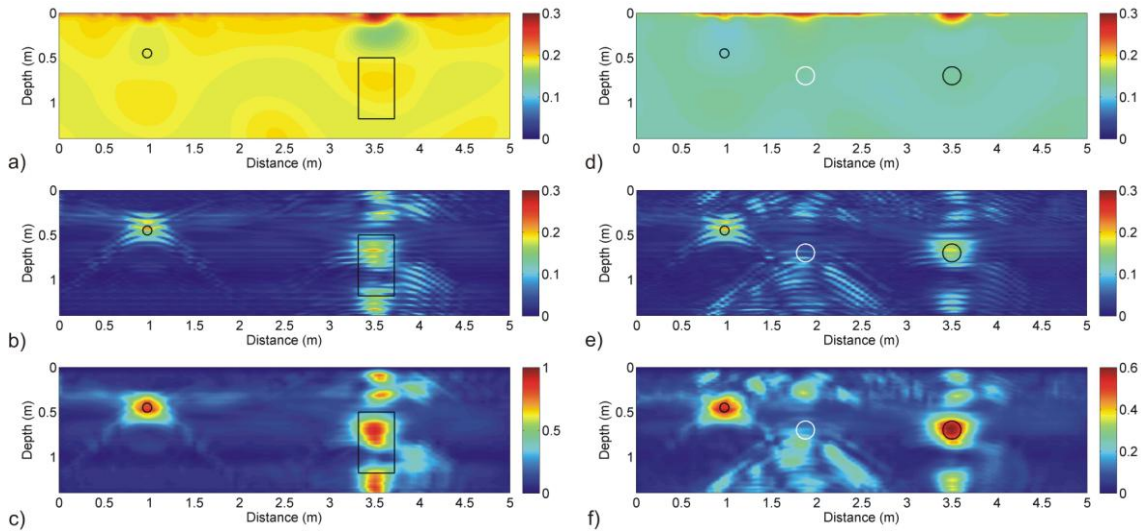

**Figure 9.** TR results for Models 1 and 2, applied to the GPR data with mean background removal. a) and d) Mode 1. b) and e) Mode 2. c) and f) Mode X12.

TR results based on eigenvalues background removal are presented in Figure 10. The pattern of the normalized TR resembles that of a wavefront focusing on the target locations. Mode 2 (Figures 10b and 10e) processing data shows a different shape anomaly but position and depth of the targets remain discernible. Mode X12 (Figures 10c and 10f) maintains the general features seen in Mode 2 but with high amplitude values for the targets, including the geologic anomaly.

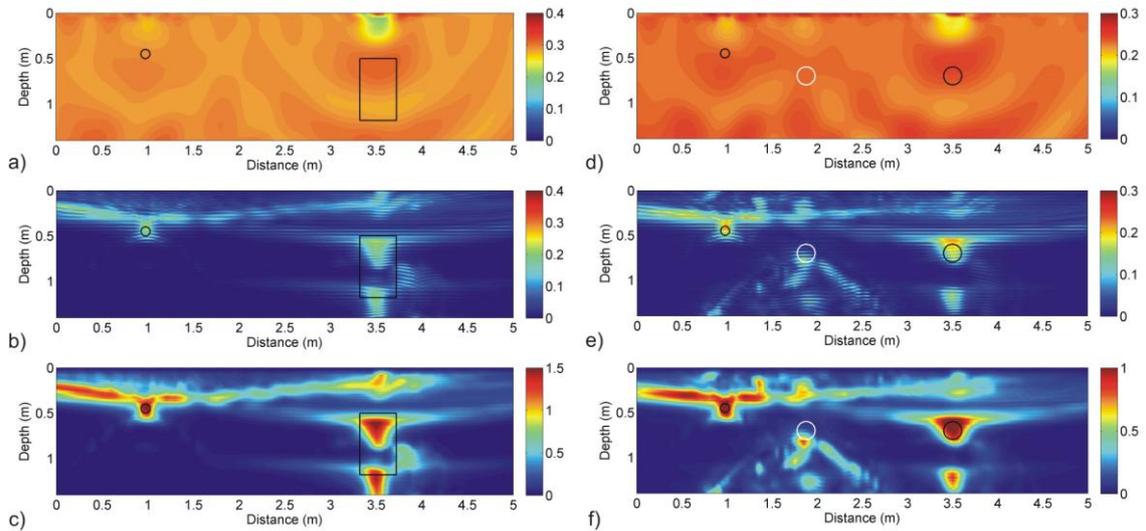

**Figure 10.** TR results for Models 1 and 2, applied to the GPR data with eigenvalues background removal. a) and d) Mode 1. b) and e) Mode 2. c) and f) Mode X12.

Using the sliding-window (Figure 11), Mode 1 shows a different pattern for the anomalies in 11a and 11d. In contrast, Mode 2 (11b and 11e) is able to better delineate the position and depth of the targets. The boundaries of the rectangular anomaly can also be determined with greater precision. Mode X12 (11c and 11f) again reduces spurious artifacts and further improve the results.

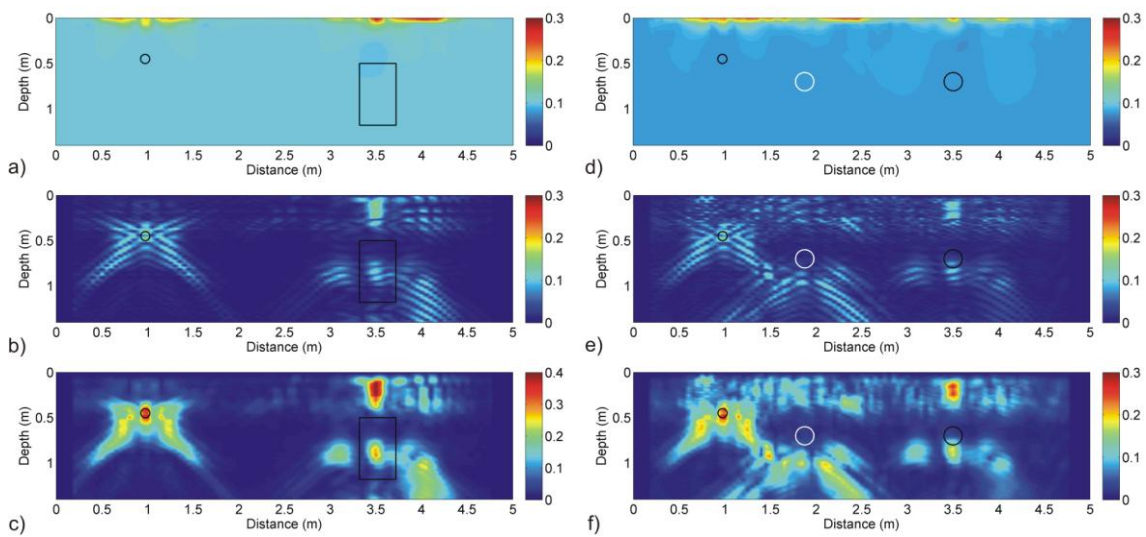

**Figure 11.** TR results for Models 1 and 2, applied to the GPR data with sliding-window processed data. a) and d) Mode 1. b) and e) Mode 2. c) and f) Mode X12.

Applying the noise-robust spatial differentiator, Mode 1 shown in Figures 12a and 12d show a different pattern for the anomalies, with a highly resolved vertical null (as expected) at the target position but again with no depth information. Mode 2 (12b and 12e), on the other hand yields depth information together with positional information along the track. In this case, the limits of the rectangular anomaly are visible, similarly to the sliding-window technique considered before. Mode X12 (12c and 12f) preserves the boundaries of the vertical metallic drum and estimates well both location and depth of the other targets.

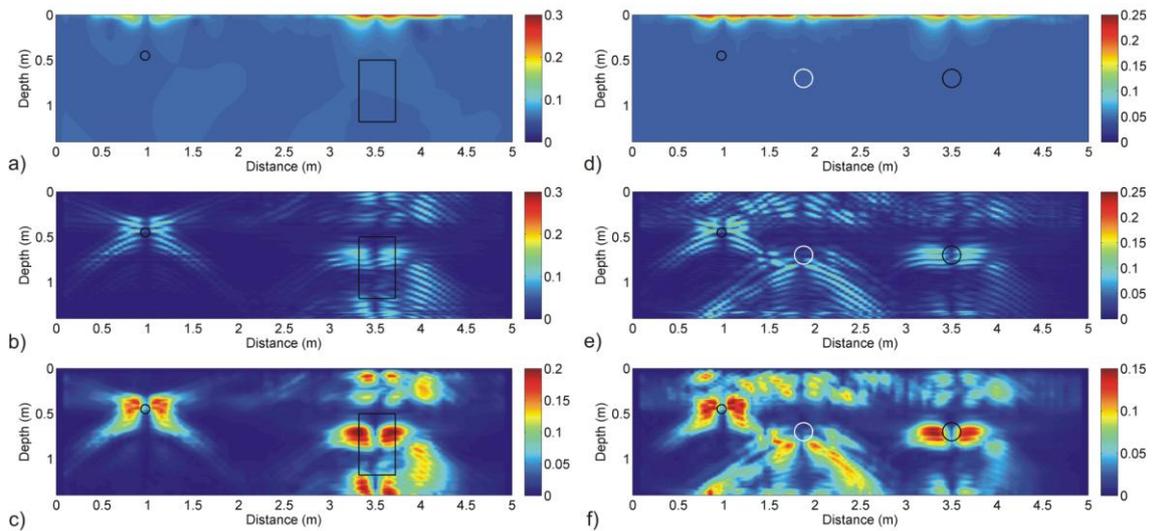

**Figure 12.** TR results for Models 1 and 2, applied to the GPR data with along-track spatial derivative processing. a) and d) Mode 1. b) and e) Mode 2. c) and f) Mode X12.

## 6. Field measurement results

GPR acquisitions were made with a GSSI (Geophysical Survey System, Inc.) system operating at central frequency of 200 MHz. The targets are buried plastic and metallic pipes. Figure 13a show a picture of the data acquisition campaign.

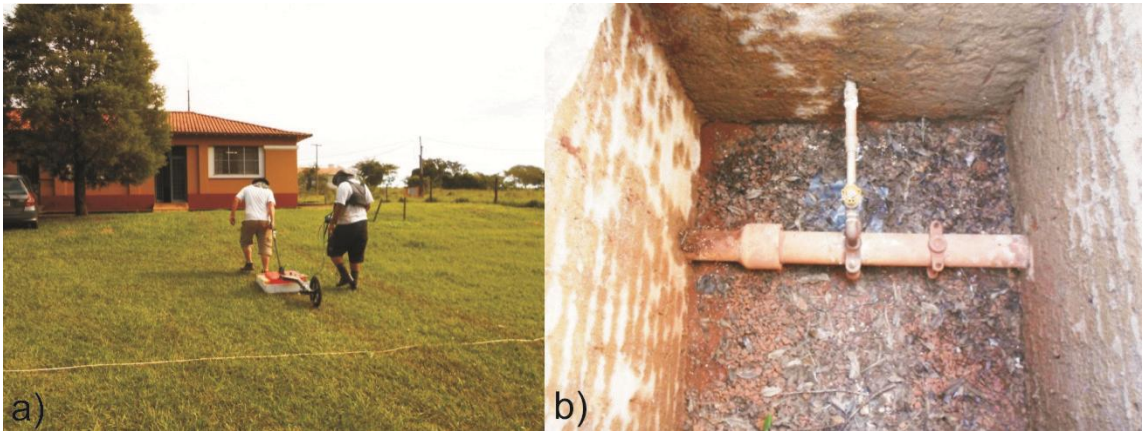

**Figure 13.** Acquisition with 200 MHz GPR system.

The first example of GPR field data is a 10 m long radargram obtained in an area with a sandy-clay soil with $\varepsilon_r \simeq 18$ and $\sigma \simeq 7$ mS/m (estimated values). The target is a buried fiber pipe, shown in Figure 13b, positioned at 7.5 m and at a depth of approximately 0.6 m. Figure 14 shows the raw radargram obtained as well as the subsequent results following the four pre-processing techniques discussed before. The target hyperbola is visible in all data sets. The results from the application of the TR-based processing are depicted on Figure 15, where the first column corresponds to Mode 1, second column Mode 2 and third column to Mode X12. Each row corresponds to a different pre-processing technique. All pre-processing techniques provide good target resolution along track. The result using eigenvalues background removal show the best results for track resolution using Mode 1. It also provided good depth resolution using Mode 2. The results with the sliding-window data and the along-scan derivative data also provided good discrimination results (the target location is indicated by the null on the trace). Mode X12 again provides ringing-noise reduction in all cases. In particular, Mode X12 results based on the eigenvalues yield very precise focusing at the exact position of the target.

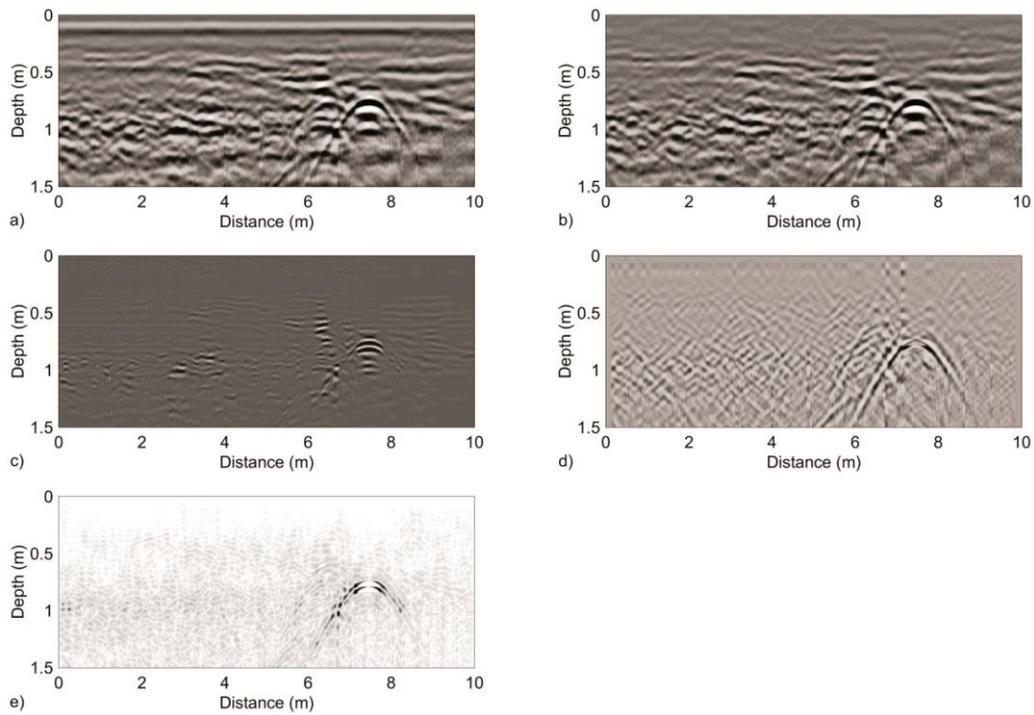

**Figure 14.** GPR field measurement data corresponding to a fiber pipe buried positioned at 7.5 m along track and at a depth of 0.6 m in a sand-clay soil. a) Raw data. b) Data with mean background removal. c) Data with eigenvalues background removal. d) Data with sliding-window data processing. e) Data with along-track spatial derivative processing.

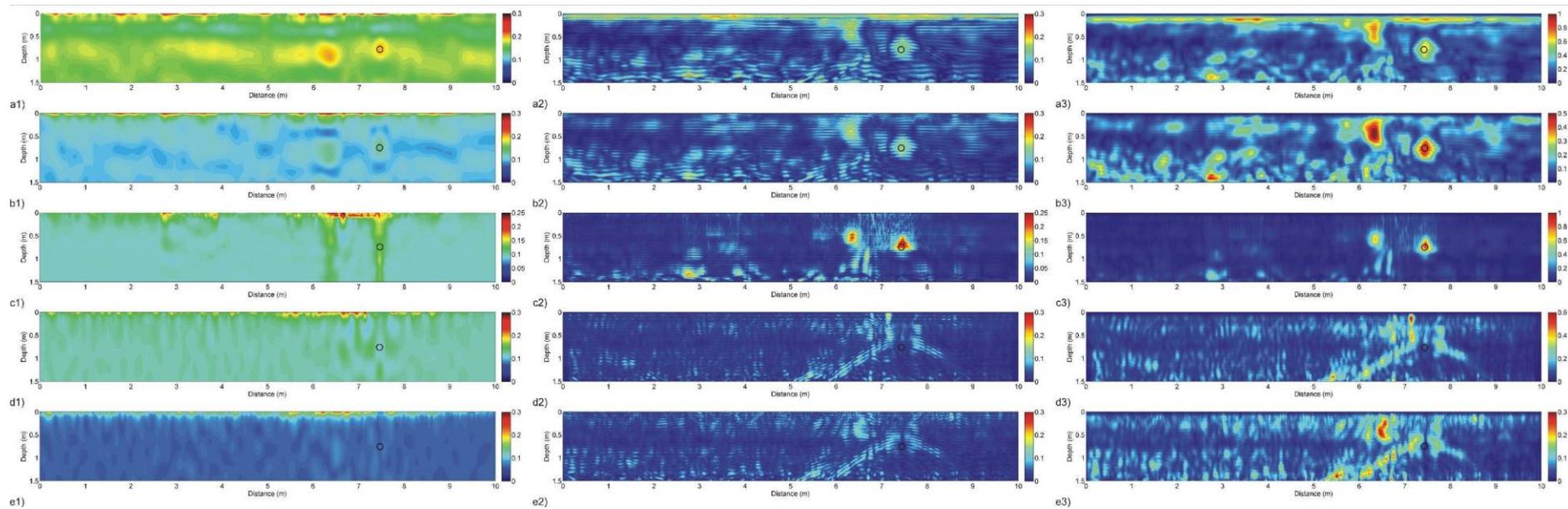

**Figure 15.** TR result applied to GPR field data of Fig. 14. Raw data: a1) Mode 1, a2) Mode 2 and a3) Mode X12. Mean background removal data: b1) Mode 1, b2) Mode 2 and b3) Mode X12. Eigenvalues background removal data: c1) Mode 1, c2) Mode 2 and c3) Mode X12. Sliding-window data: d1) Mode 1, d2) Mode 2 and d3) Mode X12. Along-track derivative data: e1) Mode 1, e2) Mode 2 and e3) Mode X12.

The second example is shown in Figure 16 and consists of two plastic pipes for water distribution buried in a clay soil environment with high loss positioned at 6 and 11 m along track, and both at 0.25 m of depth. This example corresponds to data with low SNR and, for this reason, a more challenging target hyperbole determination. By applying the pre-processing methods described above to the radargrams, the background noise is reduced somewhat. TR-based results are presented in Figure 17 with good detection performance, especially for Mode 2 and Mode X12, but also exhibiting higher clutter due to geological anomalies and noise.

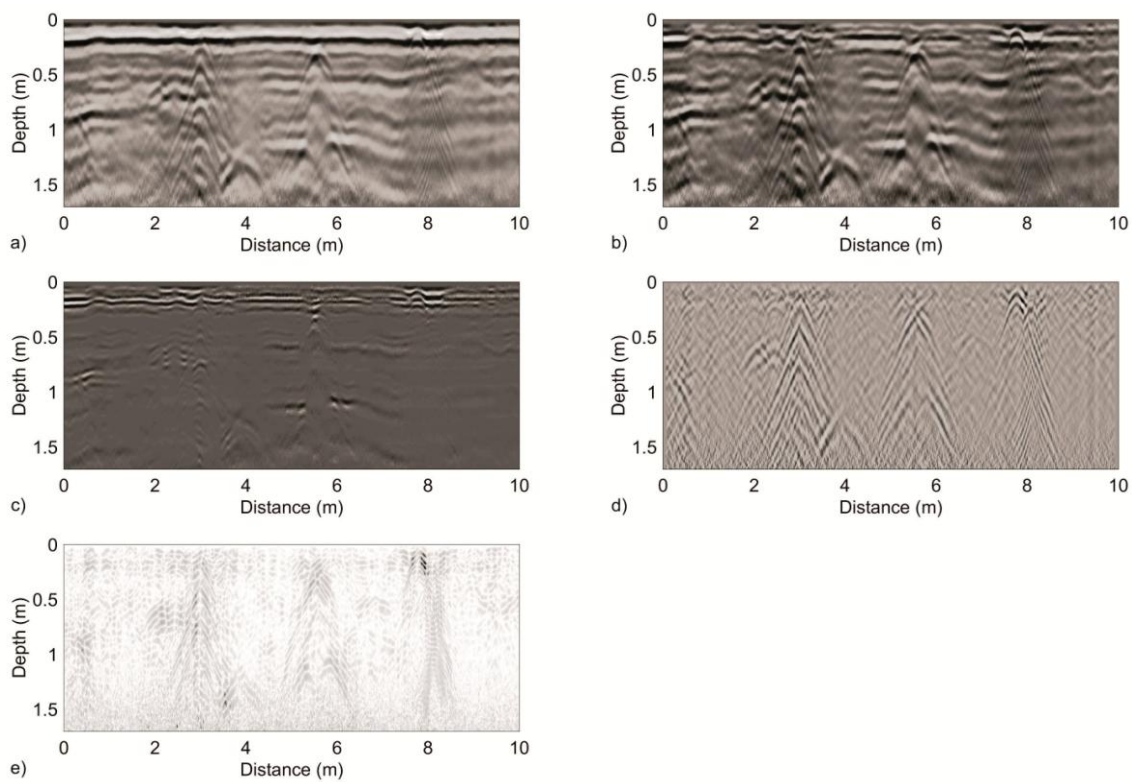

**Figure 16.** GPR field measurement data corresponding to two plastic pipes buried in a clay soil environment at 6 and 11 m along track, and both at a depth of 0.25 m. a) Raw data. b) Data with mean background removal. c) Data with eigenvalues background removal. d) Data with sliding-window data processing. e) Data with along-track spatial derivative processing.

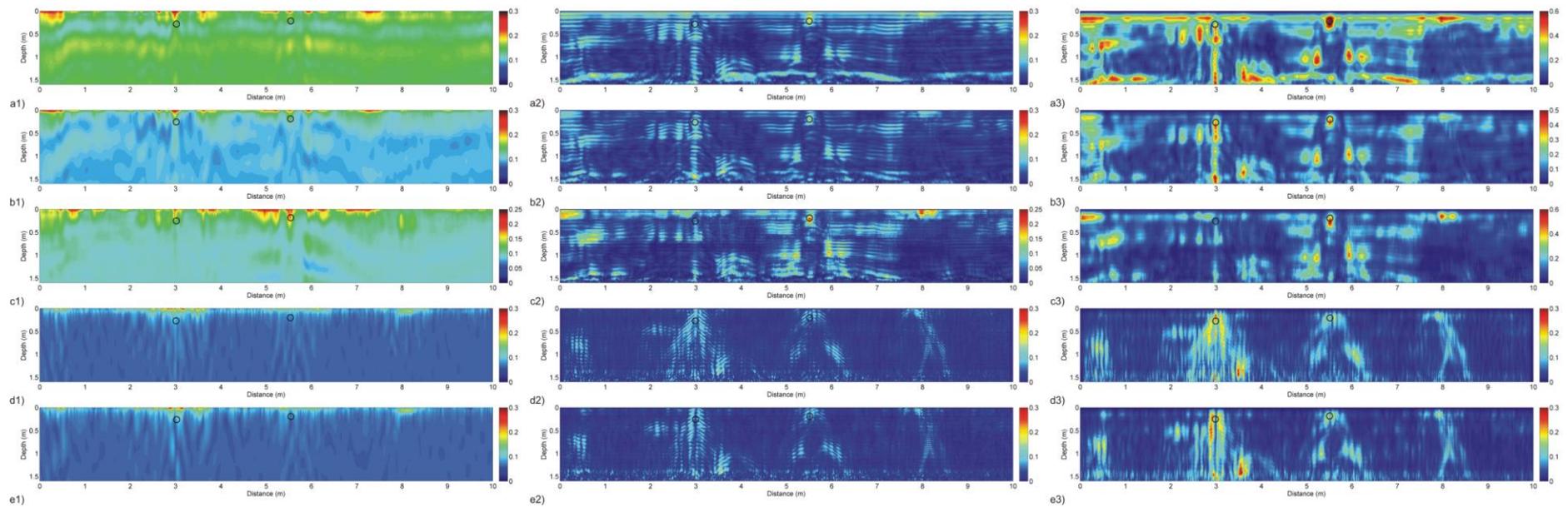

**Figure 17.** TR result applied to GPR field data shown in Fig. 16. Raw data: a1) Mode 1, a2) Mode 2 and a3) Mode X12. Mean background removal data: b1) Mode 1, b2) Mode 2 and b3) Mode X12. Eigenvalues background removal data: c1) Mode 1, c2) Mode 2 and c3) Mode X12. Sliding-window data: d1) Mode 1, d2) Mode 2 and d3) Mode X12. Along-track derivative data: e1) Mode 1, e2) Mode 2 and e3) Mode X12.

The last field data example is shown in Figure 18 and corresponds to two metallic pipes of water distribution in a clay soil. The positions of the pipes are 0.45 m and 0.93 m along track, and they are located at depths of 0.40 m and 0.50 m, respectively. The results from TR processing in raw and in conjunction with the four pre-processing techniques are presented in Figure 19. It is seen that although Mode 1 does not show good results, using Mode 2 and Mode X12 it is possible to discern the pipes position (track and depth) as well as their approximate diameters.

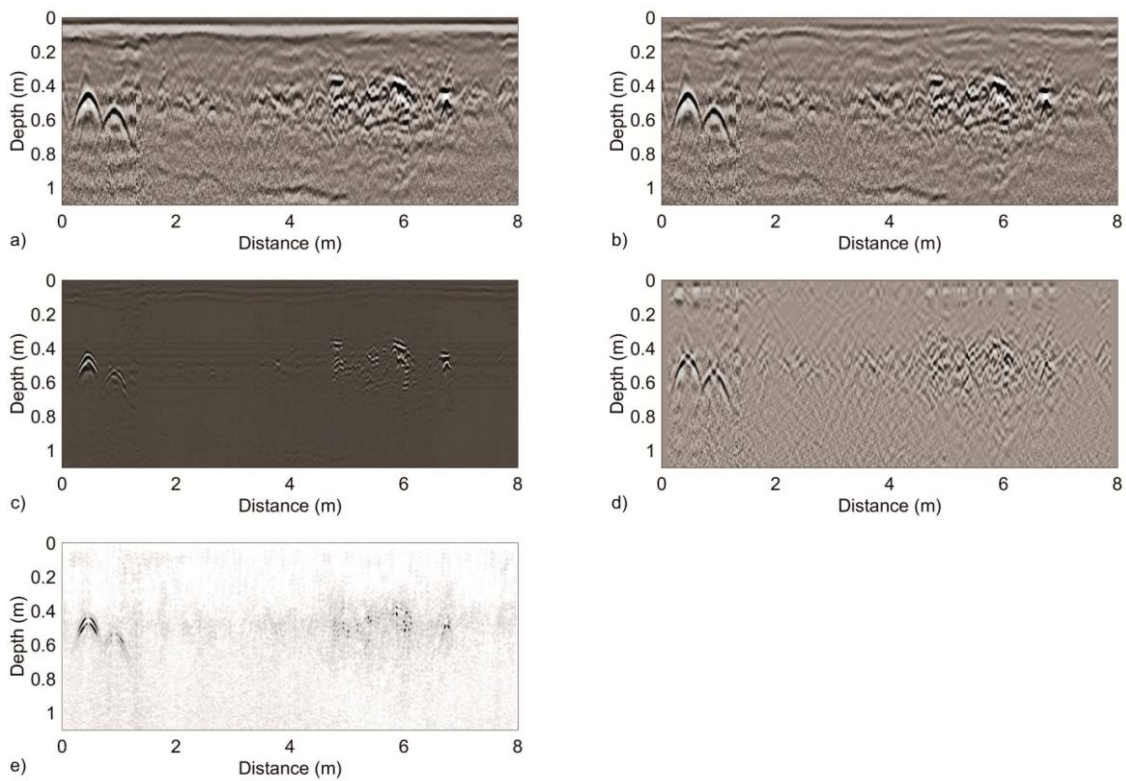

**Figure 18.** GPR field measurement data corresponding to two metallic pipes buried in a clay soil environment at 0.45 m and 0.93 m along track, and at a depth of 0.40 m and 0.50 m respectively. a) Raw data. b) Data with mean background removal. c) Data with eigenvalues background removal. d) Data with sliding-window data processing. e) Data with along-track spatial derivative processing.

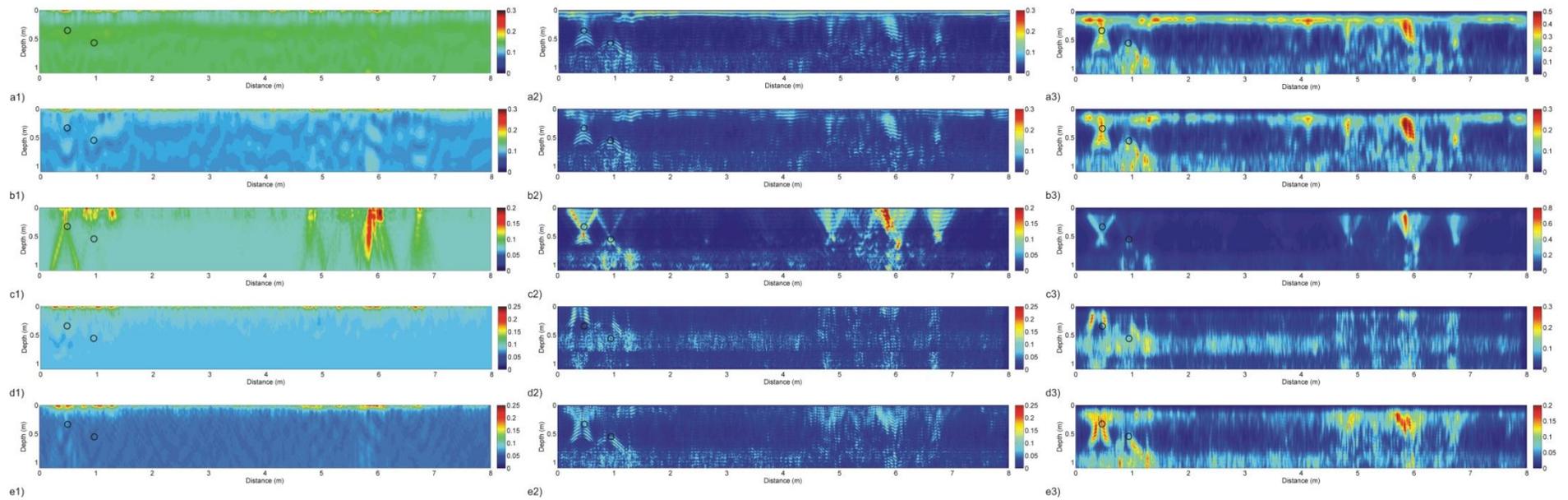

**Figure 19.** TR result applied to GPR field data of Fig. 18. Raw data: a1) Mode 1, a2) Mode 2 and a3) Mode X12. Mean background removal data: b1) Mode 1, b2) Mode 2 and b3) Mode X12. Eigenvalues background removal data: c1) Mode 1, c2) Mode 2 and c3) Mode X12. Sliding-window data: d1) Mode 1, d2) Mode 2 and d3) Mode X12. Along-track derivative data: e1) Mode 1, e2) Mode 2 and e3) Mode X12.

## 7. Conclusions

This paper considered the application of the time-reversal (TR) technique in conjunction with different pre-processing algorithms to synthetic and field GPR data. Synthetic models simulate different targets on specific environments and allow for careful control of the scattering conditions. The field data consisted of a fiber pipe buried on a sand-clay soil, and plastic and metallic water pipes buried on clay soil. The standard deviation of the raw 3D TR data and pre-processed matrices was computed based on two different data arrangements, denoted as Mode 1 and Mode 2, as well as their cross-correlation, denoted Mode X12. In general, TR results obtained were satisfactory for both synthetic and field data, and showed good potential for improving targets detection and for differentiating discrete target and geological reflections. The standard deviation and subsequent cross-correlation of TR-processed GPR data can be seen as a valuable tool for use on the analysis of whole-TR data sets from GPR campaigns. Both Mode 1 and Mode 2 provided along track targets location estimates. Under most conditions, Mode 2 provided good target depth resolution as well. Mode X12 is able to suppress some spurious artifacts of Mode 1 and Mode 2, thus aiding data interpretation. Among all data pre-processing considered here, mean and eigenvalues background removal have yielded better results, as seen in particular from the field case studies. It remains to be seen if these observations remain valid for other types of targets and for different soil conditions.

**Acknowledgements**

This work was supported in part by the Brazilian Agency National Council for Scientific and Technological Development - CNPq under Grant: 211340/2014-6. The authors also thank the Department of Geophysics of IAG/USP for providing the GPR equipment, J. L. Porsani, E. R. Almeida, C. A. Bortolozo, M. C. Stangari and E. C. Santos for technical support. The authors acknowledge the editor and reviewers for their valuable suggestions.